\documentclass[10pt, conference]{IEEEtran}  

\IEEEoverridecommandlockouts                              

\renewcommand{\baselinestretch}{.99}

\usepackage{algorithm,amssymb,url}
\usepackage{epsfig}
\usepackage{amsmath}
\usepackage{amsthm} 
\usepackage{amssymb}
\usepackage{graphicx}

\usepackage[noend]{algpseudocode}
\usepackage{epsfig}
\usepackage{amsmath,amsthm} %
\usepackage{amssymb}  %
\usepackage[english]{babel}
\usepackage{cite}
\usepackage{verbatim}
\usepackage{url}
\usepackage{float}
\usepackage{url}
\usepackage{wrapfig}
\usepackage{graphicx}
\usepackage[scriptsize]{caption}
\usepackage{multirow,subfig}

\makeatletter
\def\BState{\State\hskip-\ALG@thistlm}
\makeatother

\newcommand{\mc}{\mathcal}

\newtheorem{theorem}{Theorem}

\newtheorem{remark}{Remark}
\newtheorem{corollary}{Corollary}

\newenvironment{sistema}%
{\left\lbrace\begin{array}{@{}l@{}}}%
{\end{array}\right.}

\renewcommand{\baselinestretch}{.985} 

\begin{document}

\title{Adapting Caching to Audience Retention Rate: \\
Which Video Chunk to Store?}

\author{Lorenzo Maggi$^\star$\thanks{$^\star$Mathematical and Algorithmic Sciences Lab, France Research Center, Huawei Technologies Co. Ltd.}, Lazaros Gkatzikis$^\star$, Georgios Paschos$^\star$, and J\'er\'emie Leguay$^\star$}

\maketitle
\thispagestyle{empty}


\begin{abstract}
Rarely do users watch online contents entirely. We study how to take this into account to improve the performance of cache systems for video-on-demand and video-sharing platforms in terms of traffic reduction on the core network. We exploit the notion of ``Audience retention rate'', introduced by mainstream online content platforms and measuring the popularity of different parts of the same video content. 
We first characterize the performance limits of a cache able to store parts of videos, when the popularity and the audience retention rate of each video are available to the cache manager. We then relax the assumption of known popularity and we propose a LRU (Least Recently Used) cache replacement policy that operates on the first chunks of each video. We characterize its performance by extending the well-known Che's approximation to this case. We prove that, by refining the chunk granularity, the chunk-LRU policy increases its performance. It is shown numerically that even for a small number of chunks ($N=20$), the gains of chunk-LRU are still significant in comparison to standard LRU policy that caches entire files, and they are almost optimal.
\end{abstract}



\section{Introduction}

Content Distribution Networks (CDN) and Video on Demand applications use
network caches to store the most popular contents near the user and reduce backhaul bandwidth expenditure. 
The future projections for the cost of memory and bandwidth promote the use of caching to satisfy the ever-increasing network traffic \cite{roberts2013exploring}.
Since the bandwidth saving potential of caching is restricted by the number of files that fit in the cache (the cache capacity), it is interesting to maximize the caching effectiveness under such a constraint.
Here we consider the use of \emph{partial caching}, a technique according to which we may cache specific parts of files, instead of whole ones.

\begin{figure}[!t]
\centering
\includegraphics[width=\columnwidth]{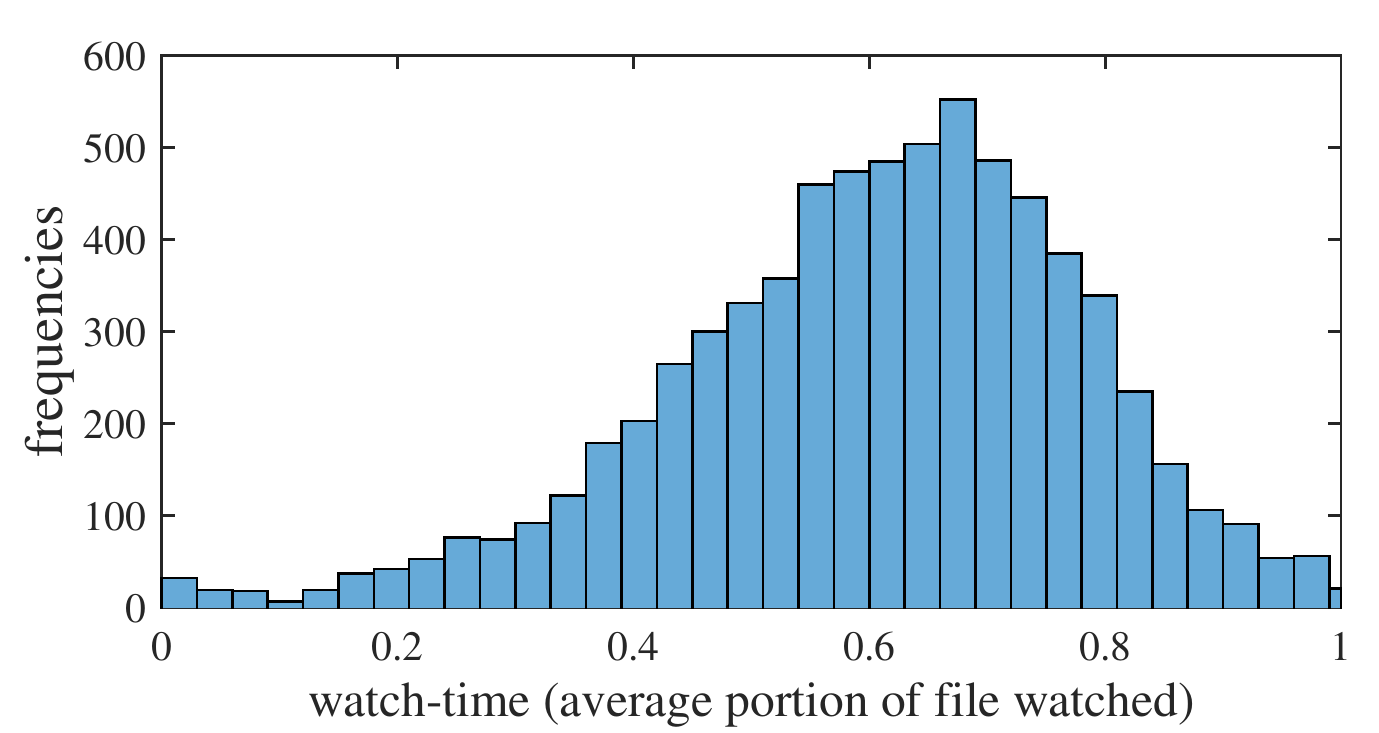}
\caption{Histogram of  watch-time in YouTube (based on a data sample of $7000$ video files from \cite{zeni2013youstatanalyzer}). On average $60$\% of a file is watched.} \label{fig:histogram}
\end{figure}

We focus on video files which represent a significant fraction of the global Internet traffic ($64$\% according to \cite{cisco_VNI}).
Videos are the most representative example of contents that are only partially retrieved, since specific parts of a video are viewed more  than others. 
Typically, the average user will ``crawl'' several video files before watching one in its entirety. 
The above imply that most of the times it is not needed to cache the entire video. 
Indeed Fig.~\ref{fig:histogram} shows the video watch-time from a trace of $7000$ YouTube videos.  The histogram emphasizes the fact that the vast majority of files is only partially watched, and motivates the design of caching algorithms that avoid caching rarely accessed video parts, e.g. the tail.  


Optimization of caching is often based on file popularity. 
Storing the most popular files results in more \emph{cache hits}, which decreases the impact on the traffic on the core network.
Nevertheless, not all the parts of a file are equally popular \cite{hwang12}.  
Hence, a natural generalization of ``store the most popular files'' is to split the video files into chunks and ``store the most popular chunks'' instead. To differentiate the popularity of each video chunk we use the metric
of the \emph{audience retention rate} \cite{retentionrate}, which measures the popularity of different parts of the same file. It has many advantages: it is file specific, it is available in most content distribution platforms, e.g., YouTube \cite{retentionrate}, and it evolves very slowly over time, which facilitates its easy estimation\footnote{The quasi-static nature of audience retention rate relates to file particularities, e.g. a movie may become uninteresting towards the end.}. The latter is not generally true for chunk popularity which are affected by the time-varying popularity of the corresponding file.

In this paper we establish a link between the audience retention rate and the efficiency of partial caching.
Our approach is based on decomposing popularity into video popularity and
video retention rate. More specifically, we address the following questions: $i$) 
\textit{ How much bandwidth could we save via partial caching of video content} and $ii$) \textit {Is this gain achievable by practical caching algorithms?}     





\subsection{Related Work}
Partial caching techniques were first reported in the context of proxy caching, where it was proposed to store the file headers to improve latency performance \cite{Sen99}.
To capture both latency and bandwidth improvements, \cite{wu04} splits the files into segments of exponentially increasing size. 
More generally, it is possible to cache specific chunks in order to capture the different popularity of sections within a file (a.k.a. internal popularity) \cite{hwang12,wang2015optimal}.

Intuitively, extreme chunking (e.g. at byte level) offers finer granularity and potentially leads to the optimal caching performance. 
However, tracking popularity at such fine granularity is impractical and leads to  algorithms of prohibitively high complexity \cite{yu2006dynamic}. 
A series of works suggest to split each file into a small number of chunks and treat each chunk independently \cite{agrawal14,wu04}.
Alternatively, it is proposed to model internal popularity as a parametric $k$-transformed Zipf distribution \cite{yu2006dynamic, lim14}. Knowing the distribution type, simplifies the estimation task
but still requires parameter estimations individually for each file. Deducing the optimal size and number of chunks is not straightforward. It was recently shown that restricting to $n$ homogeneous chunks incurs a loss which is bounded by O($n^{-2}$) \cite{wang2015optimal}. 
Alternative heuristic approaches suggest that only a specific segment of each file should be cached and dynamically adjust its size. For instance, \cite{chen05}  proposes a segmentation scheme where initially the whole object is cached but the segment size is gradually set equal to its estimated average watch-time. 
Similar adaptive strategies have been also considered for peer-to-peer networks \cite{hefeeda08}, where  starting from a small segment, the portion to be cached is  increased according to the number of requests and watch-time. 
The caching of several segments of each file was proposed in \cite{devi12}, since users may be interested only in specific, non-contiguous parts of files. In this case the segment size  has to be selected accordingly.

In this paper we prove that the performance of partial caching indeed improves when the file is split into chunks. We develop an analytical framework for LRU performance under partial caching and we use it to show that the performance gains of partial caching remain significant even for a small number of chunks. Up to the authors' knowledge, there are no studies assessing analytically the actual performance of such cache management strategies and their inherent performance limits under the partial viewing assumption. 




\subsection{Main contributions}

We first investigate a trace of YouTube data \cite{zeni2013youstatanalyzer} and conclude that partial caching has a great potential to improve performance, mainly because: (i) the average video watch-time is no more than $70\%$, and (ii) the larger the video is the less its average watch-time. Motivated by this, in Section~\ref{sec:upperbound} we present an analysis of traffic bandwidth reduction which is based on the audience retention rate.  
Combining the theoretical analysis with the YouTube data, we show that in realistic settings the traffic reduction of partial caching over traditional caching may reach up to $50\%$. 

The above analysis compares the performance limits of the two caching approaches assuming known popularity and retention rates.
Therefore, it is also interesting to investigate the bandwidth benefits from partial caching in a more realistic setting. In Section~\ref{sec:LRUchunk} we design a class of practical chunk-LRU (Least Recently Used) policies, which split files into different chunks and always drop (i.e., never cache) the last chunk at the tail of files. Chunk-LRU policies harness the realistic gain of partial caching due to video watch-time. Moreover we gain intuition into designing optimal chunking and we show that the maximum performance can be approached with a small number of chunks of equal size.

Our main technical contributions to the literature are:
\begin{itemize}
\item We formulate the traffic reduction optimization problem and provide a waterfilling algorithm to solve it efficiently. For the special case where users watch each video continuously until they abandon it, we derive the optimal waterfilling partial allocation in closed form. It consists of caching a compact interval $[0,\nu]$ of the file where $\nu$ is given in closed form.
\item We propose a novel chunk-LRU algorithm that splits each file in $N+1$ chunks where the last one is never cached.
\item We build an analytical framework to analyze the chunk-LRU performance under partial viewing, subject to Che's approximation for LRU performance, \cite{che2002hierarchical}.  
\item We provide a sufficient condition for retention rates such that sub-splitting chunks is always beneficial.
\item We characterize the optimal performance of chunk-LRU as a simple optimization problem over the tail drop factor and with infinitesimal chunking.
\end{itemize}


\section{YouTube Video Watch-time} \label{sec:motivation}

In this section we examine YouTube access traces\footnote{The dataset is publicly available and was crawled using the YouTube Data API in 2013. It contains  information about $7000$ files, including daily views, watch-time, duration, genre and title of each file.} \cite{zeni2013youstatanalyzer} in order to analyze the average video watch-time, which is the portion ($\in[0;1]$) of each file watched by the users. Watch-times are crucial for caching: using partial caching we may avoid to cache rarely watched parts of videos and use the freed cache space to store more files.



\begin{figure}[!htb]
\centering
\includegraphics[scale=.35]{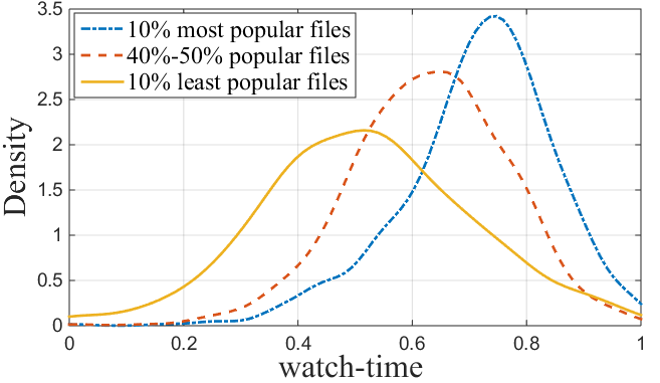}
\caption{Watch-time distribution for different classes of video popularity. The average watch-time of a video increases with its popularity.}
\label{fig:density}
\end{figure}

Since most strategies try to cache the most popular files, first we investigate the relationship between average watch-time and file popularity. We classify videos into $10$ groups according to their average daily views. Fig.~\ref{fig:density} depicts the estimated probability density function of watch-time for three representative groups, the 10\% most popular videos, the 10\% least popular, and the intermediate ones. Interestingly, we observe that \emph{the more popular a video is, the higher the average watch-time}.
However, even for the most popular ones, on average only $72\%$ of each video is watched, which leaves room for caching optimization.      

\begin{figure}[!tb]
\centering
\includegraphics[scale=.5]{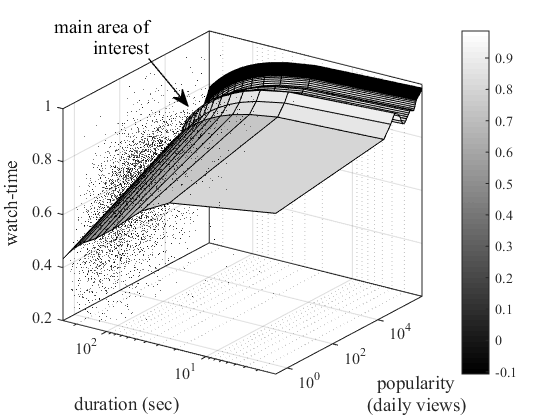}
\caption{Average watch-time is increasing with the popularity of files, but steeply decreasing with its duration.}
\label{fig:3d_pop_dur}
\end{figure}

Next, we investigate the relationship between watch-time and video duration. 
The latter is a critical parameter for caching due to the cache capacity constraint which eventually determines caching performance. 
If longer videos are only partially watched, avoiding to cache their unwatched parts will yield a greater benefit.
In Fig.~\ref{fig:3d_pop_dur} we depict with dots the YouTube data for the $20\%$  most popular files. In order to identify how the watch-time is affected by the video duration and its popularity, we use  locally weighted polynomial regression \cite{cleveland1979robust} to fit a smoothed surface to the corresponding data. Notice that the most beneficial regime for caching purposes corresponds to the upper left corner of the plot, namely highly popular videos of large size. We observe that in this region the average watch-time is around $0.7$. In addition, independently of the video popularity, \emph{watch-time decreases rapidly with video duration}.   

We then group the available data to $10$ classes 
according to their popularity and duration ($\gtrless 200$ sec).  We depict the details of the derived classes in Table~\ref{tab:class}, namely for each class we depict the average watch-time, the fraction of videos belonging to this class and its average duration in seconds. We observe that the \emph{large and popular videos amount to a non-negligible percentage  of $5\%$}. In addition, the average watch-time of large files is significantly smaller than that of smaller ones. 
To precisely evaluate the impact of watch-time to caching, we use these data in the subsequent Sections~\ref{sec:upperbound},\ref{sec:LRUchunk} to quantify the theoretical maximum and the practically feasible caching performance.

 \begin{table*}
\begin{center} 
 \begin{tabular}{| l| |c | c | c || c | c |c | } 
 \hline
    \multirow{2}{*}{Popularity \textbackslash Duration } &     \multicolumn{3}{|c||}{Small} & \multicolumn{3}{c|}{Large} \\
 			&  Av. watch-time 	& Fraction of population 	&  Av. Duration (sec)	& Av. watch-time 	& Fraction of population 	& Av. Duration (sec)\\
  \hline  Lowest 	&    0.52   		& 	0.179		 & 	81	 &	0.37		& 0.020		& 220 \\
  \hline  Low		&   0.6   		&	 0.162	 	& 	112 	&	0.47		& 0.036		& 220 \\
   \hline Medium	&   0.64		&	 0.153	 	& 	128 	&	0.57		& 0.045		&223 \\
    \hline  High	&    0.67  		&  	0.152  	& 	130 	&	0.60		& 0.047 		& 222 \\
    \hline  Highest	&     0.72		&    	0.145	 	& 	124 	&	0.65		& \textbf{0.053}	& 235 \\
 \hline
 \end{tabular}
 \caption{The characteristics of each class of videos. These data will be used to derive realistic and class-specific retention rates for our numerical evaluation.} \label{tab:class}
 \end{center} 
 \end{table*}

\section{System Model} \label{sec:sys_model}

We consider a communication system where users download video contents from the network. Let $\mc M=\{1,\dots,M\}$ be the video content (or simply, video) catalog. Each video $i\in \mc M$ is of size $S_i$ bytes. Content requests are generated using the well-known Independent Reference Model (IRM) \cite{fricker2012versatile} according to which the requests 
for the videos $\mc M$ are independent of each other. We call $p_i$ the probability that video $i$ is requested, given that a video request has arrived. Equivalently, the sequence of video requests can be thought of as $M$ independent homogeneous Poisson processes with intensity rate proportional to the probability vector $\{p_i\}_i$. For convenience of notation, we assume that the probabilities are in decreasing order, i.e., $p_1\ge p_2 \ge \dots \ge p_M$.

\begin{figure}[!htb]
\captionsetup{belowskip=-0pt,aboveskip=4pt}
\centering
\includegraphics[scale=.55]{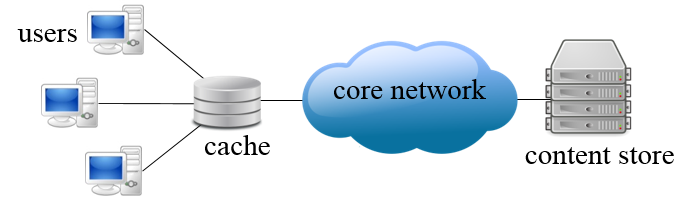}
\caption{System model}
\label{fig:core_network_better}
\end{figure}

One cache of size $C$ bytes is deployed in the network.\footnote{Our analysis can be extended to a cache hierarchy by letting $p_i$ express the probability that a request for video $i$ is missed by the caches at all the child nodes \cite{roberts2013exploring}.} 
 Whenever a requested video  is found in the cache, the cache itself can directly serve the user. Otherwise, the video needs to be retrieved through the core network, which provides access to a central video content store containing the entire video catalog, see Fig.~\ref{fig:core_network_better}. 
Hence, good caching performance has a profound impact on the traffic reduction on the core network. The goal of this paper is to determine the extra bandwidth benefits that may be gained by exploiting the fact that videos are rarely watched entirely.

\begin{table}[t]
\centering
{\begin{tabular}{|p{.08\textwidth}| p{.37\textwidth}|}
\hline
$\mc M$ & : video catalog of cardinality $|\mc M|=M$ \\ 
$C$ & : cache size \\ 
$p_i$ & : popularity of video $i$ \\ 
$R_i(\tau)$ & : audience retention rate of video $i$ \\
$\pi_i(\tau)$ & : viewing abandonment p.d.f. of video $i$ \\ 
$S_i$ & : size of video $i$ \\ 
$B_s(\boldsymbol{Y})$ & : traffic bandwidth on the core network when the portion $Y_i$ of video $i$ is statically stored in the cache (see Eq. \eqref{eq:Bs}) \\
$N\!+\!1$ & : number of chunks for chunk-LRU \\
$\underline{B}$ & : minimum core network traffic achieved by optimal partial caching \\
$[x_{k-1},x_k]$ & : $k$-th chunk of a video \\
$\mathbf{x}$ & : collection of chunks \\
$\nu$ & : tail drop factor for chunk-LRU; the last chunk $[\nu;1]$ is never stored in the cache \\ 
$h_{k,i}$ & : hit rate of the $k$-th chunk of video $i$ \\
$t_C$ & : characteristic time for chunk-LRU \\
$B_{\mathrm{cLRU}}(\mathbf{x},\nu)$ & : traffic on core network with chunk-LRU (see Eq. \eqref{eq:BcLRU}) subject to the chunking $\mathbf x$ and a tail drop factor $\nu$ \\ 
$\underline{B}_{\mathrm{cLRU}}$ & : optimal traffic performance for chunk-LRU (see Eq. \eqref{eq:mainoptLRU}) \\
\hline
\end{tabular}}
\caption{Table of notation symbols}\label{tab:sym}
\end{table}

\subsection{Viewing Behavior Model: Audience Retention Rate} \label{sec:downloadbehav}

To mathematically analyze the impact of watch-time, we introduce the central notion of audience retention rate $R_i(\tau)$. According to Youtube's definition, the audience retention rate $R_i(\tau)$ measures the percentage of users that are still watching video $i$ at the corresponding (normalized) instant $\tau$, out of the overall number of views \cite{retentionrate}.
As we will see, in our analysis the retention rate has a prominent role in determining the caching performance. 

Typically a user may watch video $i$ from instant $a_i(1)$ up to $b_i(1)$, then she possibly skips to $a_i(2)$ and watches until $b_i(2)$, and so forth\footnote{We remark that such intervals may also overlap, i.e., a user may rewind the video and watch a part of it multiple times. We assume that, if this occurs, then the user can directly retrieve the video portion that she has already watched from her terminal's cache.}. The watched part $W_i$, which equals the minimum portion of video $i$ that the user needs to download, is the union of all watch intervals $j$:
\[
W_i = \, \cup_j [a_i(j);b_i(j)].
\]
We call $|W_i|$ the \emph{watch-time} of user watching video $i$. 
For ease of notation we consider $a_i,b_i\in[0;1]$ as portions of the whole video duration. The ``\emph{audience retention rate}\footnote{Our definition is in accordance with the definition of audience retention (or ``engagement'') rate by Wistia.com \cite{wistiaeng}. Youtube's audience retention rate \cite{retentionrate} actually counts the video rewinds as multiple views inside the same videos.}'' function $R_i(\tau)$ can be then formally defined as the probability that a user has watched the (normalized) instant $\tau$ of the video, i.e.,
\[
R_i(\tau) = \, \Pr \left( \tau\in W_i \right), \qquad \tau\in [0;1].
\]
Alternatively, we may think of $R_i(\tau)$ as the fraction of users that watch the (normalized) instant $\tau$ of the video $i$.

We remark that, thanks to the definition of $R_i$, we can easily evaluate the average watch-time for video $i$ as $\int_0^1 R_i(\tau)d\tau$.

Next we devise a realistic and more specific viewing behavior model and we derive its relationship to audience retention rate.


\subsubsection{Viewing Abandonment Model} \label{sec:viewaband}

This is a special instance of the viewing model presented above. It assumes that users always start watching each video $i$ from its beginning, and they abandon it after a random time portion $b_i\in[0;1]$. Hence, in this case the watched part $W_i$ takes on the simple form $W_i = \, [0;b_i]$, thus $b_i$ equals the watch-time. We call $\pi_i(.)$ the probability density distribution of the abandonment time variable $b_i$. The relationship between the abandonment distribution $\pi_i$ and the audience retention rate $R_i$ is described by the expression:
\begin{align}
R_i(\tau) = & \, 1 - \int_0^{\tau} \pi_i(t)dt. \label{eq:Riaband}
\end{align}
Hence, in this case the audience retention rate $R_i(\tau)$ measures the \emph{fraction of users with watch-time higher than $\tau$} for the particular video $i$. We first observe from \eqref{eq:Riaband} that $R_i$ is inherently non-increasing, with $R_i(0)=1$. We also remark that, under the viewing abandonment assumption, the audience retention rate $R_i$ uniquely describes the random watch behavior $[0;b_i]$ of user via $\pi_i$. This observation does not hold though for the general case described in Section \ref{sec:downloadbehav}, where the same retention rate $R_i$ may result from an arbitrary distribution of watch behaviors.

\begin{figure}[!htb]
\captionsetup{belowskip=-0pt,aboveskip=4pt}
\centering
\includegraphics[scale=.6]{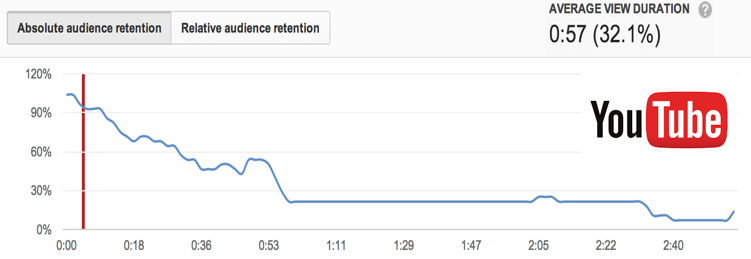}
\caption{Instance of audience retention rate from YouTube.}
\label{retention_youtube}
\end{figure}

In order to come up with a realistic audience retention rate function from the estimated parameters in Tab. \ref{tab:class} for our numerical investigations in Sections \ref{sec:perfevalupp},\ref{sec:sim} we assume that the viewing abandonment model holds. 

\section{Performance Limits of Partial Caching} \label{sec:upperbound}

This section analyzes the performance limits of partial caching in the context of audience retention rate.
Our performance metric is core network traffic and we tackle the off-line problem of finding the optimal \emph{static} (partial) file cache allocation\footnote{We remark that in our analysis of the optimal traffic bandwidth $B(\boldsymbol Y^*)$ we assumed that the videos $\boldsymbol Y^*$ are already present in the cache and we did not take into account the traffic needed to fill the cache. If we wish to incorporate this aspect, we could say that $B(\boldsymbol Y^*)$ is the expected traffic achieved asymptotically over a number of requests tending to infinity.}. In particular, we will compare the maximum network traffic saved by caching entire videos versus caching arbitrary portions of each of those. In both cases it is idealistically assumed that the video popularity distribution $\{p_i\}_{i\in\mc M}$ and the audience retention rate functions $\{R_i\}_{i\in\mc M}$ are perfectly known to the cache manager. This analysis serves as an upper bound for any cache management strategy with more limited information, as the one devised in Section \ref{sec:LRUchunk}. 

Let us first formalize our problem. We define the  \emph{partial allocation} $Y_i\subseteq [0;1]$ of video $i$ to be the collection of (possibly) non-adjacent bytes, that are selected to be \emph{permanently} stored in the cache. 
Subject to a partial allocation $Y_i$, any requests for the remaining portions 
$[0;1]\setminus Y_i$ need to be served by the origin video store. Due to the specific retention rate for this video, this happens with probability $\int_{[0;1]\setminus Y_i} R_i(\tau) d\tau$. Therefore, under a partial allocation vector $\boldsymbol Y$, we may express the expected traffic on the core network per request $B(\boldsymbol Y)$ as 
\begin{equation} \label{eq:Bs}
B(\boldsymbol Y) = \sum_{i\in\mc M} S_i p_i \int_{[0;1]\setminus Y_i} R_i(\tau) d\tau.
\end{equation}

Considering the video size $S_i$ and cache size $C$, a partial allocation vector $\boldsymbol Y$ is feasible whenever
$\sum_{i\in\mc M} S_i \int_{Y_i} 1 dx = \, C$.
Our goal is to select a feasible vector $\boldsymbol Y$ that minimizes the incurred traffic $B_s(\boldsymbol Y)$, i.e.,
\begin{align}
\boldsymbol{Y}^* = & \, \underset{\boldsymbol Y}{\operatorname{argmin}} \, B(\boldsymbol{Y}) \label{eq:uppbound1} \\
\mathrm{s.t.} & \, \begin{sistema}
\sum_{i\in\mc M} S_i \int_{Y_i} 1 dx = \, C \\
Y_i\subseteq [0;1]
\end{sistema} \notag
\end{align}
If users always watch the whole video, i.e., $R_i(\tau)=1$ for all $\tau\in [0;1]$ and $i\in\mc M$, 
then the optimization \eqref{eq:uppbound1} takes a simple form which is solved by the well-known \emph{store the most popular videos} policy. In this case, we would choose to fully store, $Y_i=[0;1]$, the videos of highest $p_i$ up to the cache capacity and no portion of the rest, i.e. $Y_i=\emptyset$ otherwise.
As indicated by the previous section however, in reality this is not the case, hence we expect $\boldsymbol{Y}^*$ to bring certain improvement, that we evaluate in Section \ref{sec:perfevalupp}. 

Technically speaking, if we lift any assumption on the shape of the audience retention rate, the best cache allocation should intuitively prescribe to partition all videos at the finest granularity (at the byte level, say), order them according to their popularity, and fill the cache with the \emph{most popular bytes}. We now provide an equivalent waterfilling characterization of the optimal partial video allocation $\boldsymbol{Y}^*$ to solve this problem. The main advantage of this formulation lies in the fact that it leads to an efficient algorithm to compute $\boldsymbol{Y}^*$, that we present at the end of the section.

\begin{theorem} \label{theo:optimalwater}
The optimal partial video allocation $\boldsymbol{Y}^*$ can be expressed as
\begin{equation} \label{eq:uppboundtheo}
Y_i^*(\mu) = \, \{\tau: \ p_i R_i(\tau)\ge \mu \} \quad \forall\, i\in\mc M,
\end{equation}
where $\mu$ is such that $\sum_{i\in\mc M} S_i |Y_i^*(\mu)| = \, C$, where $|.|$ is the size\footnote{formally defined as the Lebesgue measure} of a subset of $[0;1]$.
\end{theorem}

Informally speaking, the water level $\mu$ determines a popularity threshold above which a byte of any video deserves to be stored in the cache.

\subsection{Viewing Abandonment Model}

In the special case of viewing abandonment model, we already observed that the audience retention rate $R_i$ is non-increasing for all $i\in\mc M$. This allows us to specialize our result in Theorem \ref{theo:optimalwater} as follows.

\begin{corollary} \label{prop:downlwater}
Consider the viewing abandonment model with strictly decreasing $R_i$, for all $i\in\mc M$. The optimal video allocations writes $\boldsymbol{Y}^*=[0;\eta^*_i]$ for all $i\in\mc M$, where
\begin{equation} \label{eq:cor1}
\begin{sistema}
\eta_i^*(\mu) = \, \begin{sistema}
1 \quad \mathrm{if} \ p_i R_i(1)\ge \mu \quad (\mu\ge 0)\\
0 \quad \mathrm{if} \ p_i \le \mu \\
R_i^{-1}(\mu/p_i) \quad \mathrm{otherwise} 
\end{sistema} \\
\sum_{i\in\mc M} S_i \eta_i^*(\mu) = \, C.
\end{sistema}
\end{equation}
\end{corollary}

A remarkable observation here is that optimum bandwidth performance is achieved by splitting every video in only two parts and caching the first one. We may determine the exact splits if the abandonment distribution is given. 
 For instance, if $\pi_i$ is truncated exponential one with parameter $\lambda_i$, i.e.,
\[
\pi_i(\tau) = \, \frac{\lambda_i}{1-e^{-\lambda_i}} \, e^{-\lambda_i \tau}, \qquad \tau\in[0;1],
\]
then the following holds.

\begin{corollary} \label{cor:downlexpwater}
Under the exponential viewing abandonment model the optimal video allocations writes $\boldsymbol{Y}^*=[0;\eta^*_i]$ for all $i\in\mc M$, where
\begin{equation}
\begin{sistema}
\eta_i^*(\mu)= \left[-\frac{1}{\lambda_i} \ln\left( \frac{\mu}{p_i}(1-e^{-\lambda_i}) + e^{-\lambda_i}\right) \right]^+ \!\!\!\!, \ \quad (\mu\ge 0)\\
\sum_{m=1}^M S_i \eta_i^*(\mu) = \, C.
\end{sistema}
\end{equation}
\end{corollary}

\subsection{Computation of Optimal Performance}

To solve \eqref{eq:uppbound1}, we observe that it can be expressed as a separable convex optimization problem with linear and box constraints. If we further assume that the functions $R_i$ do not have any plateau, then the objective function becomes strictly convex, thus we can adapt the algorithm presented in (Section 7.2, \cite{stefanov2013separable}) to our scope in order to efficiently compute the optimal cache partial video allocation $\boldsymbol{Y}^*$. We present below the high-level description of the algorithm. An interested reader may find in the Appendix the implementation details.


\begin{small}

\noindent\rule{3.5in}{0.01in}

\noindent\textbf{Waterfilling algorithm}

\vspace{-0.09in}
\noindent\rule{3.5in}{0.01in}

\noindent Set $k:=0$. Set $\mc M^{(0)}:=\mc M$.\\
\noindent \textbf{while} $\mc M^{(k)}\ne \emptyset$

\begin{itemize}

\item refine the search of the set of indices $\mc M^{(k)}$ in correspondence to which the optimal solution is deemed to be in the interior of the box constraint

\item if the approximated solution for video $i\in\mc M^{(k)}$ falls beyond the box $[0;1]$, it is rounded to the nearest boundary; it is now optimal and discarded from $\mc M^{(k)}$

\item set $k:=k+1$
\end{itemize}

\noindent \textbf{end}

\noindent\rule{3.5in}{0.01in}\vspace{-8pt}
\end{small}

\subsection{Performance Evaluation with Real Data} \label{sec:perfevalupp}

\begin{figure}[hbtp]
\captionsetup{belowskip=-0pt,aboveskip=4pt}
\centering
\includegraphics[scale=.5]{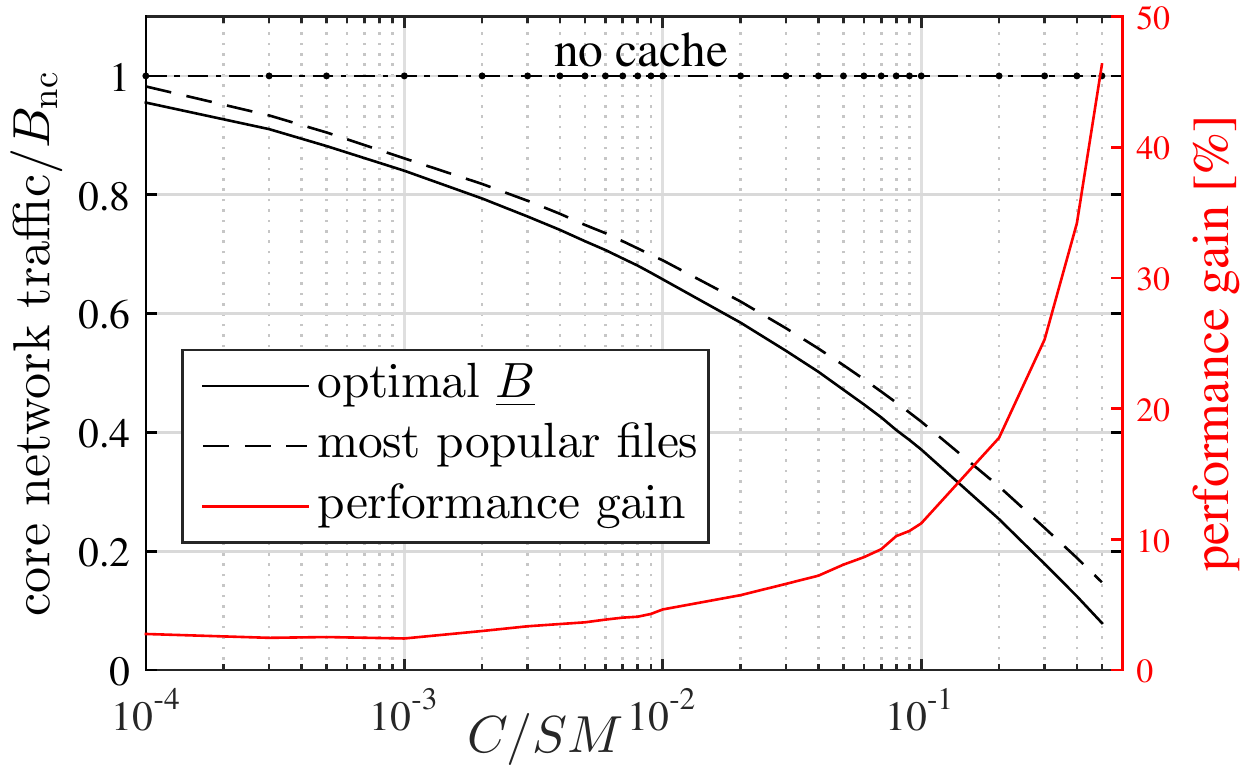}
\caption{Core traffic generated by the optimal partial caching strategy in a realistic scenario vs. the traffic produced by storing the most popular videos in their entirety. We show in red the resulting performance gain by using the first strategy. We utilized the parameters obtained via real data shown in Tab. \ref{tab:class}. The video popularity distribution follows a Zipf law with parameter 0.8 \cite{fricker2012versatile}. $S$ is denoted as the average video size.}
\label{fig:opt_vs_MostPop}
\end{figure}

In order to evaluate the performance of the optimal partial allocation in a realistic scenario we utilize the average watch-time parameters shown in Tab. \ref{tab:class}. In Fig. \ref{fig:opt_vs_MostPop} we compare the core network traffic $\underline{B}=B_s(\boldsymbol{Y}^*)$ generated by the optimal partial caching strategy with the one produced by the most natural strategy prescribing to store the most popular videos in their entirety. We observe that remarkable gains from partial caching are achieved for cache size ratios higher than $10^{-2}$ of the total catalog size, which we typically find in current CDN scenarios.

\begin{figure}[!htb]
\captionsetup{belowskip=-0pt,aboveskip=4pt}
\centering
\includegraphics[width=.4\textwidth]{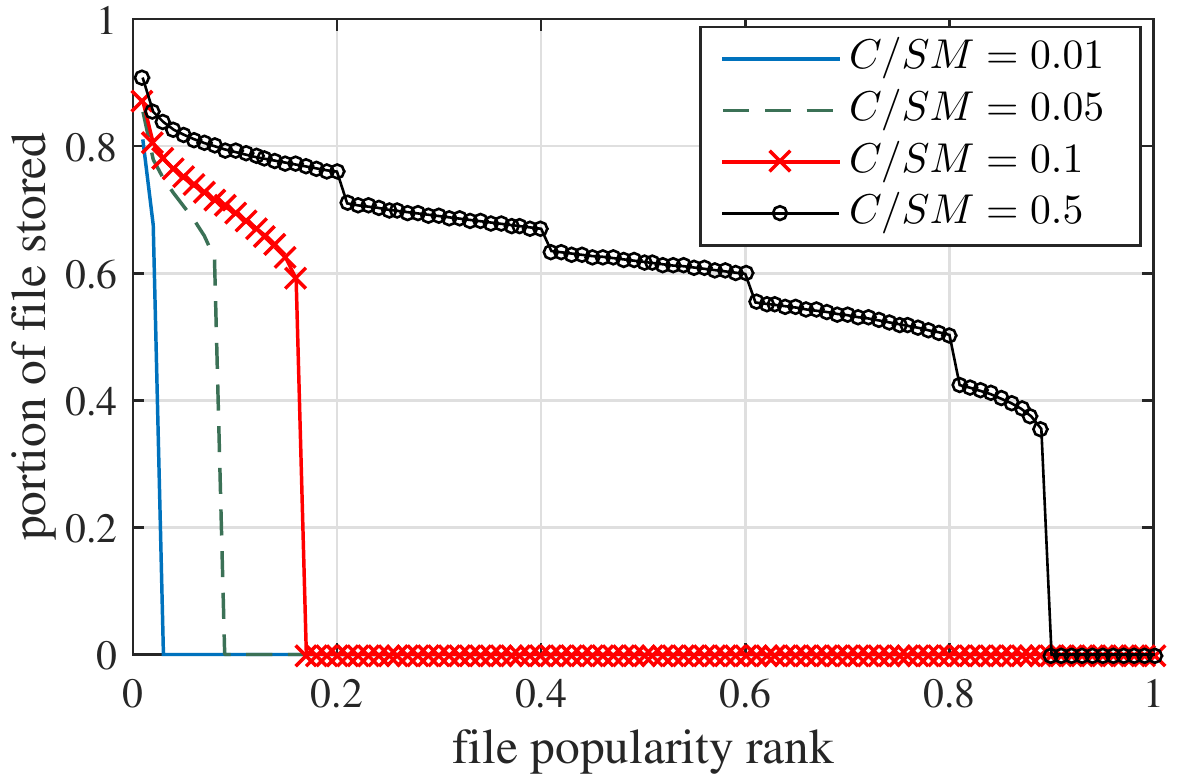}
\caption{Optimal portion of videos that should be stored according to the same optimal caching strategy in Fig. \ref{fig:opt_vs_MostPop}. Given a certain $C/SM$, the video with video popularity $x$ should be stored from its beginning up to portion $y$.}
\label{fig:stored_portion}
\end{figure}

We then show in Fig. \ref{fig:stored_portion} the optimal portion of videos that should be stored according to the same optimal caching strategy, for different values of the cache size. 
Interestingly, only very popular videos are stored in their entirety, even for large cache sizes. 


We finally remark that in this paper we normalize all the core network traffic figures with respect to the minimum bandwidth per video request $B_{\mathrm{nc}}$ required to serve the users when no cache is deployed in the system, which equals 
\begin{equation} \label{eq:Breq}
B_{\mathrm{nc}} = \, \sum_{i=1}^M S_i p_i \int_0^1 R_i(\tau) d\tau.
\end{equation}

\section{A Practical Chunk-LRU Scheme for Decreasing Retention Rates} \label{sec:LRUchunk}

After analyzing the best performance that can only be achieved with full information on the system parameters, we turn to the study of a practical cache update scheme that shows good performance even when popularity $p_i$ and audience retention rate $R_i$ are unknown for each video $i$.

It is a widespread understanding that the Least Recently Used (LRU) cache replacement policy represents a good trade-off between hit-rate performance and implementation complexity in a real scenario where no statistics on video popularity are available to the cache manager. Moreover, thanks to its short memory it reacts quickly to variations in video popularity. In its simplest form though, each time a video is requested even only partially by a user and is not found in the cache, LRU would prescribe to cache it \emph{in its entirety} (and to update the LRU recency table accordingly). Since users rarely watch videos entirely, as previously observed, the standard LRU would generate extra-traffic in the core network and would waste precious cache space to store unpopular portions of files.

In order to counter this, we propose a new cache management policy that generalizes the classic LRU policy. We first suggest to split each video into $N\!+\!1$ consecutive and non-overlapping chunks. We denote by $[x_{i-1};x_i]$ the $i$-th chunk. Moreover, we argue that the last (i.e., the $(N+1)$-th) chunk of each video, which is the least popular part under the assumption of decreasing audience retention rate, should \emph{never} be stored in the cache, even if requested by a user. Intuitively, this frees up space for more popular chunks of less popular videos to be stored in the cache. We call $\nu$ the tail drop factor that pinpoints the position of the last chunk. Hence, the first $N$ chunks of each video are stored only if requested, and then evicted from the cache in an LRU fashion.
\begin{remark}
For the sake of analysis simplicity we assume that the chunk splitting $\mathbf{x},\nu$ does not depend on the identity of the file. We leave this as a future extension.
\end{remark}

\begin{figure}
\centering
\includegraphics[width=.35\textwidth]{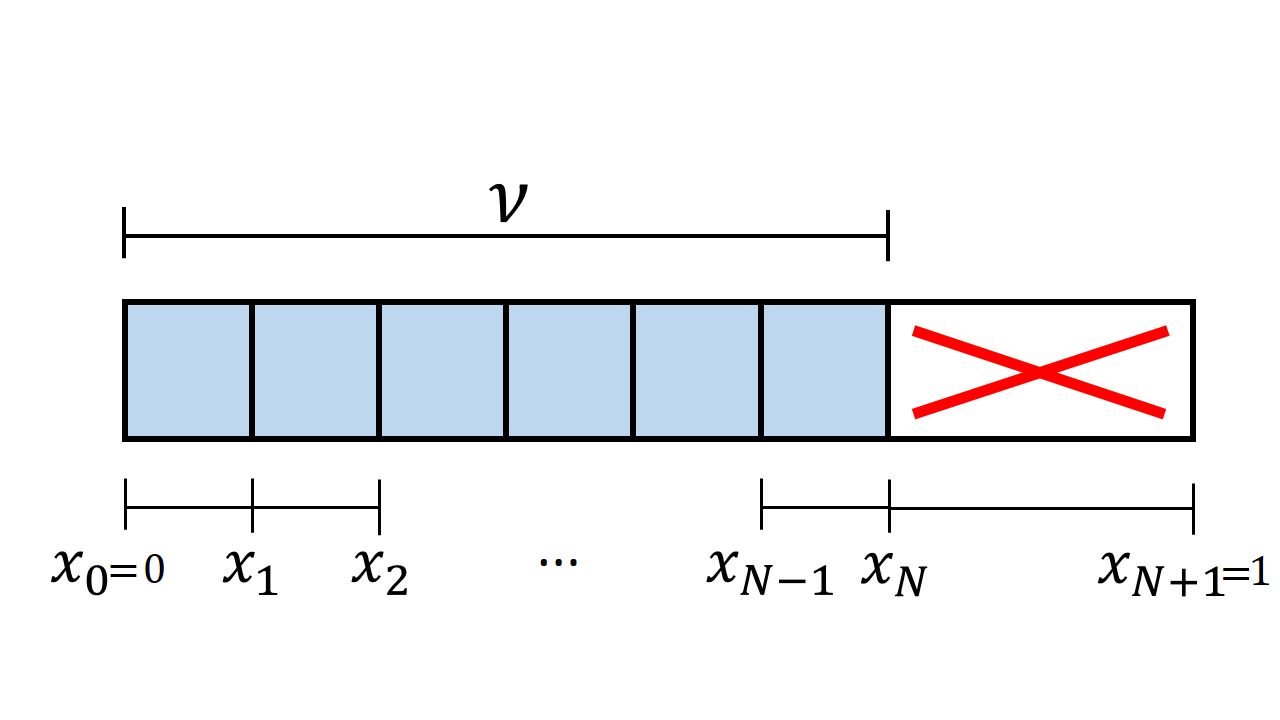}
\caption{Video split into $N\!+\!1$ chunks. Only the first $N$ are considered for chunk-LRU; the last one is never stored in the cache.}
\label{fig:chunks}
\end{figure}
Performing LRU on the first $N$ chunks presents two main benefits. On the one hand, it reduces the extra-traffic on the core network caused for the retrieval of video portions that are not requested. For instance, whenever a user watches a video from its beginning up to portion $b$, only the first $\bar k=\min_k\{ x_k\ge b \}$ chunks are downloaded. Hence, only the portion $x_k-b$ is stored in the cache without being accessed. On the other hand, we exploit the fact that the tail of a video is generally less popular than the rest \cite{yu2006dynamic}. Hence, by systematically discarding the tail of each video we avoid to evict from the cache the first chunks, which are likely to be more popular.\footnote{Additionally, although this is not the focus of this paper, performing LRU on chunks would allow to keep track of the evolution of the popularity of each chunk. Nevertheless, the resulting benefits would be minor, since the retention rate varies on a time scale much slower than the video popularity dynamics.}

We now formally describe our algorithm which uses as input the chunking of files and the tail drop factor. The impact of those parameters on actual performance is analyzed in the following subsections. 


\begin{footnotesize}

\noindent\rule{3.5in}{0.01in}

\noindent\textbf{chunk-LRU Algorithm}

\vspace{-0.09in}
\noindent\rule{3.5in}{0.01in}

\noindent\textbf{Step 1 (Initialization)}:

\begin{itemize}
\item[\textbf{1.1})] Set the tail drop factor $\nu\in(0;1]$
\item[\textbf{1.2})] Partition each video $i$ into $N\!+\!1$ chunks of the form $[x_0=0;x_1]$,$[x_1,x_2]$,$\dots$,$[x_{N-1};\nu\equiv x_N]$,$[x_N=\nu;x_{N+1}=1]$, where $x_i\in[0;1]$ (see Fig. \ref{fig:chunks})
\item[\textbf{1.3})] An initial chunk request recency vector is available
\end{itemize}

\noindent\textbf{Step 2}: A request for a packet of video $i\in\mc M$ belonging to its $k$-th chunk $[x_{k-1},x_k]$ arrives 
\begin{itemize}
\item[\textbf{2.1})] If $k=N\!+\!1$, then the request is handled by the core network and the cache is not updated (i.e., the tail is never cached)
\item[\textbf{2.2})] Else, if $1\le k\le N$, then
\begin{itemize}
\item[\textbf{2.2.1})] If the requested chunk is stored in the cache, then the cache sends the packet to the user
\item[\textbf{2.2.2})] If the requested chunk is not stored in the cache, then it is retrieved from the core network and then stored in the cache, after evicting the minimum number of least recently used chunks. Finally, the cache sends the packet to the user
\end{itemize}
\item[\textbf{2.3})] The recency vector of the chunks stored in the cache is updated in an LRU fashion
\item[\textbf{2.4})] Return to step \textbf{2})
\end{itemize}

\noindent\rule{3.5in}{0.01in}
\end{footnotesize}

\subsection{Chunk-LRU Performance under Viewing Abandonment} 

After having described our chunk-LRU algorithm, we now turn to the analysis of its performance. To this purpose, in this section we will assume that \emph{the viewing abandonment model holds}. Moreover, in order to come up with our analytical results we make the common simplifying assumption that all videos have the same size $S=S_i$. This is well justified by the fact that we can break large videos into equal size fragments, and perform chunk-LRU over the chunks of the video fragments.

We first observe that, under the viewing abandonment model (Section \ref{sec:viewaband}), the probability that the $k$-th chunk of video $i$ is requested by a user knowing that the user herself has already started watching video $i$ equals $R_i(x_{k-1}) = \int_{x_{k-1}}^1 \pi_m(\tau) d\tau$. Since the requests for video $i$ follow by assumption a Poisson process of intensity (proportional to) $p_i$, then the request process for the $k$-th chunk is also Poisson with reduced intensity $p_i R_i (x_{k-1})$. Thus, thanks to an adaptation of the popular Che's approximation \cite{che2002hierarchical} we can already compute the hit rate for a specific chunk, i.e., the probability that a chunk is found in the cache when requested.

Let us elaborate on this. Che's approximation was originally proposed in \cite{che2002hierarchical} to compute the hit rate for files whose request successions follow independent Poisson processes. It approximates the characteristic time $t_C$, measuring the time that a file spends in the cache, as a constant. When shifting the request granularity from the video to the chunk level, the independence property of request streams is unavoidably lost. Nevertheless we can still rely on the intuition that when the cache size is significantly larger than the video size the characteristic time of each chunk is approximately equal and constant, hence Che's approximation still holds, which has been shown valid in \cite{roberts2013exploring}. Therefore, the hit rate $h_{k,i}$ for the $k$-th chunk of video $i$ can be approximated as $h_{k,i} = \, 1 - e^{-p_i R_i (x_{k-1}) t_C}$, where the characteristic time $t_C$ obeys the following relation \cite{fricker2012versatile}:
\begin{equation} \label{eq:BcLRU1}
\frac{C}{S} = \, \sum_{k=1}^N \Delta x_k \sum_{i=1}^M h_{k,i},
\end{equation}
where $\Delta x_k = x_k\!-\!x_{k-1}$. Finally, the expected traffic per video request $B_{\mathrm{\mathrm{cLRU}}}$ forwarded to the core network when the chunk-LRU cache management policy is employed writes 
\begin{align}
& B_{\mathrm{\mathrm{cLRU}}}(\mathbf x,\nu) = \label{eq:BcLRU} \\
& S\sum_{i=1}^M p_i \left(\sum_{k=1}^N R_i (x_{k-1}) (1\!-\!h_{k,i}) \Delta x_k + \int_{\nu}^1 \!R_i(\tau) d\tau \right) \notag 
\end{align}
where $\mathbf x=\{x_1,\dots,x_{N-1}\}$.

\subsection{Benefits of Chunk Sub-Splitting}

We now focus on the impact of  the chunk size on chunk-LRU performance, measured as the traffic generated at the core network $B_{\mathrm{\mathrm{cLRU}}}$. Intuitively speaking, shrinking the chunk size should translate into better traffic performance, since this reduces the traffic surplus generated when users do not watch a chunk in its entirety. Nevertheless this does not prove the intuition, since modifying the chunk size also has an impact on the characteristic time $t_C$ in a non-trivial way via Eq. \eqref{eq:BcLRU1}.
 
Before stating the main result of this section, we first need to introduce some notation. Let $\underline{t}_C$ and $\overline{t}_C$ be the characteristic times when only one chunk (i.e., $[0;\nu]$) and chunks of infinitesimal size $dx$ (say, at the byte level) are employed, respectively. More formally, $\underline{t}_C$ and $\overline{t}_C$ are the unique roots of the two following equations:
\vspace{-2pt}
\begin{align*}
\frac{C}{S} = & \, \nu \sum_{i=1}^M \left( 1 - e^{-p_i \underline{t}_C} \right) \\
\frac{C}{S} = & \, \sum_{i=1}^M \int_0^{\nu} \left( 1 - e^{-p_i R_i(x) \overline{t}_C} \right) dx,
\end{align*}
\vspace{-2pt}
respectively. Moreover, we say that the chunk split $\mathbf{x}'$ is a sub-split with respect to $\mathbf{x}$ whenever $\cup_i \{ x_i \} \subset \cup_i \{ x'_i \}$. We finally observe that if $\nu=\frac{C}{MS}$ then the cache can store all the first videos up to their portion $\nu$; hence, it is reasonable to constrain $\nu$ within the interval $[\frac{C}{MS};1]$.\\
We are now ready to prove that any refinement of the chunk granularity produces a decrease in the expected traffic load on the core network.

\begin{theorem} \label{theo:infchunk}
Let $\nu\in[\frac{C}{MS};1]$ and let $\mathbf x$ be a video chunk split. Assume that 
\begin{equation} \label{eq:suffcondinf}
\frac{d}{d\tau} \sum_{i=1}^M p_i R_i(\tau) e^{-p_i R_i(\tau) t_C} \!<\! 0, \qquad  \forall\, t_C \in [\underline{t}_C;\overline{t}_C], \tau \in[0;1]
\end{equation}
Then, any video chunk sub-split $\mathbf{x}'$ outperforms $\mathbf{x}$ in terms of traffic generated on the core network, i.e., the following holds:
\[
B_{\mathrm{cLRU}} (\mathbf{x}',\nu) < \, B_{\mathrm{cLRU}} (\mathbf{x},\nu).
\]
\end{theorem}

Numerical experiments suggest that our sufficient condition \eqref{eq:suffcondinf} is very loose, and it generally holds for realistic popularity distributions and retention rates. It is not satisfied only in pathological cases where the distribution is extremely concentrated around few popular files and the cache size very small, near to the size of a single file.

\subsection{Optimal Performance of Chunk-LRU}

In this section we focus on the computation of the best performance of chunk-LRU, optimized over the chunk size and tail drop factor $\nu$. We will utilize it as a benchmark for the performance evaluation of practical chunk-LRU policies in realistic scenarios in Section \ref{sec:sim}.

In order to come up with the best performance achievable by chunk-LRU we need to find the solution of the following optimization problem:
\begin{align}
& \underline{B}_{\mathrm{cLRU}} = \, \underset{N,\mathbf{x},\nu,t_C}{\min} \, B_{\mathrm{cLRU}}(\mathbf{x},\nu) \label{eq:mainoptLRU} \\
& \mathrm{s.t.} \ 
\begin{sistema}
\frac{C}{S} = \, \sum_{k=1}^N \Delta x_k \sum_{i=1}^M 1 - e^{-p_i R_i (x_{k-1}) t_C} \\
\frac{C}{MS} \le \, \nu \le 1 \\
0=x_0\le x_1\le \dots\le x_{N-1}\le x_N=\nu. \notag
\end{sistema}
\end{align}

It follows from Theorem \ref{theo:infchunk} that, if condition \eqref{eq:suffcondinf} holds, then the bandwidth utilization of any video chunk split $\mathbf{x}$ and $\nu\in[\frac{C}{MS};1]$ is lower bounded by the performance $\underline{B}_{\mathrm{cLRU}}(\nu)$ of the infinitesimal split (say, at the byte level). This greatly simplifies the formulation of \eqref{eq:mainoptLRU} in a two-variable constrained optimization problem (see Eq. \ref{eq:infinitesimalLRU}). Below we formalize this result.

\begin{corollary} \label{cor:infintoptimal}
Assume that condition \eqref{eq:suffcondinf} holds. For any video chunk split $\mathbf{x}$ and tail drop factor $\nu$, the traffic performance $B_{\mathrm{cLRU}}(\mathbf{x},\nu)$ is lower bounded by the performance $\underline{B}_{\mathrm{cLRU}}$ of the infinitesimal chunking approach:
\[
\underline{B}_{\mathrm{cLRU}} \le \, B_{\mathrm{cLRU}}(\mathbf{x},\nu),
\]
where $\underline{B}_{\mathrm{cLRU}}$ is computed as
\begin{align}
& \underline{B}_{\mathrm{cLRU}} = \underset{\nu,t_C}{\min} \, \sum_{i=1}^M \int_0^{\nu} \!\!p_i R_i(x) e^{-p_i R_i(x) t_C} dx + \int_{\nu}^1 \!\!p_i R_i(\tau) d\tau  \notag \\
& \mathrm{s.t.} \ 
\begin{sistema}
\frac{C}{S} = \, \sum_{i=1}^M \int_0^{\nu} \left( 1-e^{-p_i R_i(x) t_C} \right) dx \\
\frac{C}{MS} \le \nu \le 1.
\end{sistema} \label{eq:infinitesimalLRU}
\end{align}
\end{corollary}

We stress the fact that $\underline{B}_{\mathrm{cLRU}}$ is the lowest core network traffic  achievable by a chunk-LRU cache management policy.

Thanks to the formulation in \eqref{eq:infinitesimalLRU} we can prove the following two results via standard Lagrangian optimization techniques. 

\begin{corollary} \label{cor:nuoptmin1}
If $R_i$ is continuous and $R_i(1)=0$ for all $i\in \mc M$ then the optimal $\nu^*<1$.
\end{corollary}

\begin{corollary} \label{cor:nuopteq1}
If $R_i(\tau)=1$ for all $\tau\in[0;1]$, $i\in \mc M$ then standard LRU (only one chunk for each video and $\nu=1$) achieves optimal performance.
\end{corollary}

The former result states that if users never watch videos in their entirety, then it is always optimal to never cache a non-negligible portion of file, i.e., $\nu^*<1$. The latter claims that, as intuition suggests, if all users watch the whole video then the best chunk-LRU policy is actually the standard LRU.

\subsection{Performance evaluations with real data} \label{sec:sim}

\begin{figure*}
\captionsetup{belowskip=-0pt,aboveskip=4pt}
\begin{minipage}[t]{0.6\textwidth}
  \centering
\includegraphics[scale=.6]{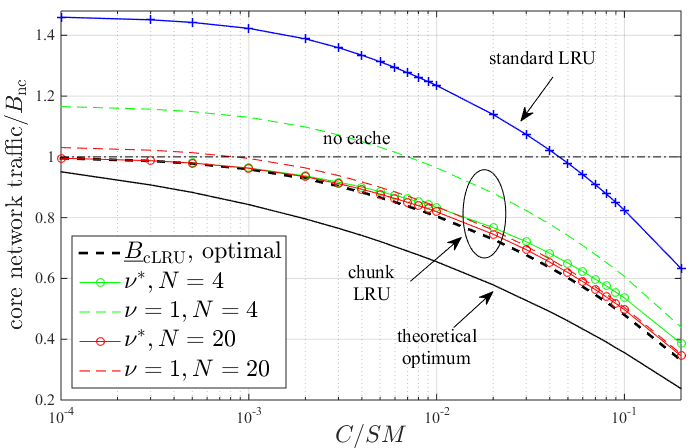}
\caption{Normalized core network traffic generated by chunk-LRU vs. the theoretical optimum $\underline{B}$ and vs. the standard LRU. The optimal $\nu^*=\nu^*(N)$ is computed for each value of $N$ and cache size $C$, as depicted in Fig. \ref{fig:nu_star_final}. We also evaluate the performance achieved when the sub-optimal value of $\nu=1$ is utilized. The video popularity distribution follows a Zipf law with parameter 0.8 \cite{fricker2012versatile}}
\label{fig:B_LRU_vs_optimal_vs_LRU_final}
  \end{minipage}
  \ \ \ \ \ \
\begin{minipage}[t]{0.4\textwidth}
\centering
\includegraphics[scale=.4]{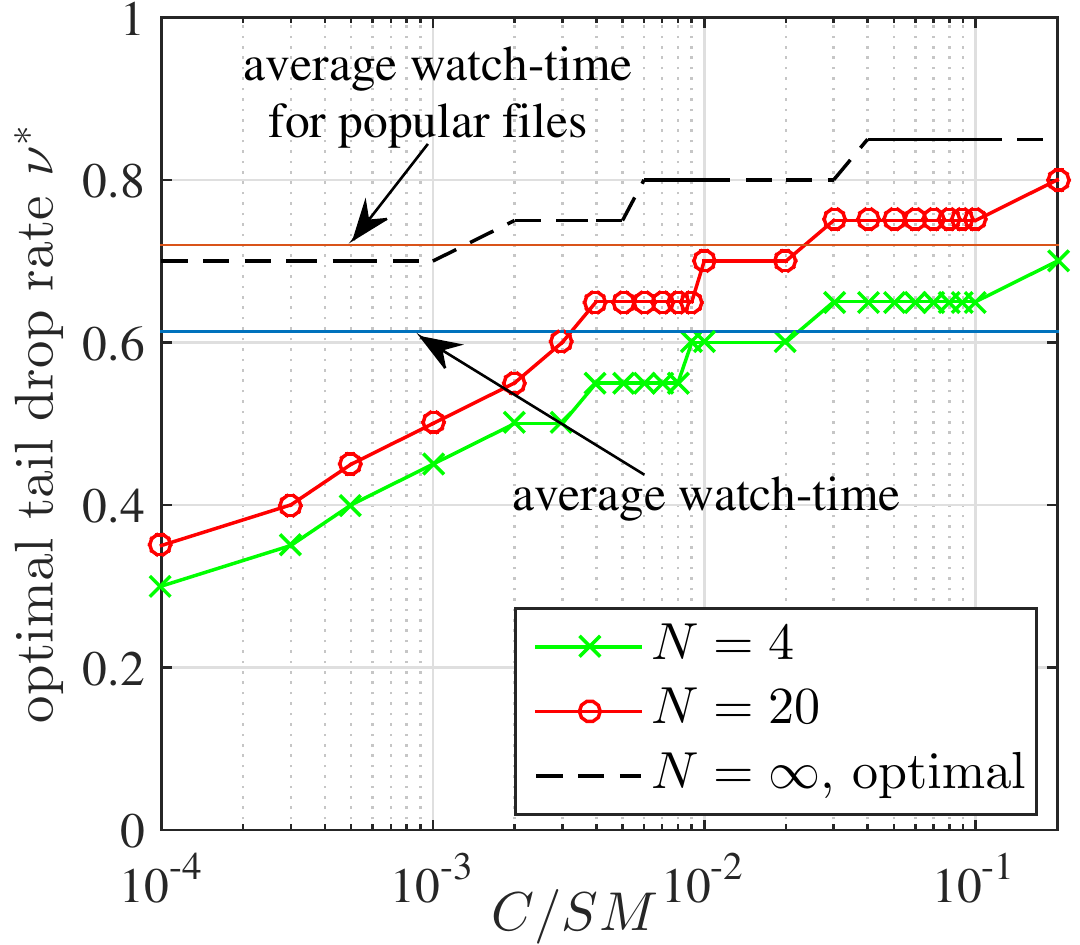}
\caption{Optimal tail drop factor $\nu^*$ for different number of chunks $N=4,20,\infty$. We notice that the optimal $\nu^*(N)$ is within a neighborhood of the average watch-time of $0.61$.}
\label{fig:nu_star_final}
\end{minipage}
\end{figure*}

In this section we numerically evaluate the traffic performance on the core network of the proposed class of chunk-LRU cache management policies. We compare them with the optimal performance $\underline{B}$ under full information that we derived in Section \ref{sec:upperbound}. We also take the performance of standard LRU as a second term of comparison. As in Section \ref{sec:upperbound}, we consider the audience retention rate scenario shown in Tab. \ref{tab:class}, estimated from a real Youtube dataset, with the only difference that the file size is supposed to be uniform. We show our results\footnote{The traffic performance is normalized w.r.t. the traffic $B_{\mathrm{nc}}$ generated when no cache is present, as in Section \ref{sec:upperbound}.\\The chunk-LRU policies have chunks with equal size.} in Fig. \ref{fig:B_LRU_vs_optimal_vs_LRU_final}. We first notice that, as hinted by Theorem \ref{theo:infchunk}, the traffic generated by chunk-LRU decreases as the number $N$ of chunks increases ($N=4,20$). The infinitesimal chunk size approach ($N=\infty$) is shown to achieve optimal performance $\underline{B}_{\mathrm{cLRU}}$, as claimed in Corollary \ref{cor:infintoptimal}. Notably, the chunk-LRU performs close to its optimal performance even with a limited number of chunks ($N=20$ or also $N=4$). Moreover, a suboptimal value of the tail drop factor $\nu=1$ still performs close to optimal for $N$ sufficiently high (see Sect. \ref{chunk_LRU_param} for further details). On the other hand, as expected, standard LRU performs poorly. In fact, the traffic generated by retrieving parts of file that are not requested by the users outweighs the obtained benefits through cache hits even for medium-size caches. This explains why the traffic generated by LRU can be even higher than the one without any cache deployed. 

The best tail drop factor $\nu^*=\nu^*(N)$ used to produce Fig. \ref{fig:B_LRU_vs_optimal_vs_LRU_final} is optimized for each value of $N$ and cache size $C$, as shown in Fig. \ref{fig:nu_star_final}. We notice that $\nu^*$ is closely related to average watch-time, since it captures the portion of files with the lowest popularity which need to be systematically discarded from the cache. For small cache sizes, simulations show that $\nu$ is lower than the watch-time: in fact, to compensate for the reduced cache size, low values of $\nu$ allow to squeeze in the cache a significant amount of different - and popular - headers of files.

\subsection{Tuning the chunk-LRU parameters}
\label{chunk_LRU_param}

Although the optimization of chunk-LRU parameters is beyond the scope of this paper, next we provide guidelines on how reasonable values could be selected. 

\paragraph{Choosing the number of chunks} Increasing the number of chunks translates into an increase of the frequency at which the cache content and the associated recency list is updated, as well as an increase of the recency table size. Therefore, the design of the optimal number $N$ of chunks in real systems should capture the trade-off between the actual performance of the policy (for which high values of $N$ are preferable, see Cor. \ref{cor:infintoptimal}) and the required processing/memory resources, increasing with $N$. Our numerical results in Fig.~\ref{fig:B_LRU_vs_optimal_vs_LRU_final} suggest that even a small number of chunks (around 4), that would result to a low complexity policy, can achieve reasonably good traffic performance.

\paragraph{Choosing the tail drop factor $\nu$} The \emph{exact} optimal value $\nu^*(N)$ can be computed by solving the problem in \eqref{eq:infinitesimalLRU} only if all the system parameters, i.e., the file popularity $p_i$ and the retention rates $R_i$, are known to the cache controller. For comparison purposes\footnote{If the full information assumption holds then using chunk-LRU would be highly suboptimal, since the theoretically optimal solution computed in Section \ref{sec:upperbound} can be actually implemented.}, we then show in Fig. \ref{fig:B_LRU_vs_optimal_vs_LRU_final} the performance achieved in the extreme case where the cache manager is \emph{agnostic} to $p_i$ and $R_i$ and the tail drop parameter $\nu$ is blindly set to $1$, i.e., no chunks are ever discarded. Remarkably, if the number of chunks is sufficiently high ($N=20$ in this case), the loss in performance incurred by such sub-optimal choice is limited: the fine granularity of chunk splitting compensates for the loss incurred by setting $\nu=1$. 
\begin{remark}
We claim that a reasonable choice of $\nu$ ($< 1$) can be still made in realistic scenarios, based on an estimation of the parameters $p_i,R_i$. First of all, indeed, the optimal $\nu^*$ is not strictly a function of the popularity of each video, but only of the rank-dependent popularity $p_i$ of the $i$-th most popular video, for each $i$. It has been shown \cite{fricker2012versatile} that such rank-dependent relation depends on the class of traffic and is slowly varying over time, hence it is easily predictable. Secondly, we argue the video retention rate functions $R_i$ vary on a much slower time scale than that of video popularity, which greatly facilitates its estimation.
\end{remark}

\section{Conclusions}

In this paper we investigated the potential of partial caching towards minimizing core network traffic. Our numerical results based on real YouTube access data reveal that big caches benefit the most from such strategies, namely up to $50\%$ over the classic approach of storing the most popular files. Interestingly, partial caching is beneficial even when the actual popularity of videos is not known. In this case, practical chunk-based LRU strategies which never cache the tail of videos were shown to perform well as long as a sufficient number of chunks is used. 

The introduction of audience retention rate in caching decisions opens up interesting research directions. Retention rate is generally available in online video distribution systems and does not evolve over time. Thus, it can be used to decompose the problems of file popularity estimation and optimal chunking without loss of optimality. In this context, the generalization of existing caching mechanisms so as to optimally exploit the benefits of partial caching is an interesting topic for future study.

\section{Appendix}

\subsection{Waterfilling Algorithm} \label{sec:appWF}

\begin{footnotesize}

\noindent\rule{3.5in}{0.01in}

\noindent\textbf{Algorithm to compute $\boldsymbol{\eta}^*$}

\vspace{-0.09in}
\noindent\rule{3.5in}{0.01in}

\noindent\textbf{Step 1 (Initialization)}: Let $k=0$, $C^{(0)}:=C$, $\mc M^{(0)}:=\mc M$, $\mc M_a^{\mu}:=\emptyset$, $\mc M_b^{\mu}:=\emptyset$. Define $\tilde{R}'_i:\mathbb R\rightarrow \mathbb R$ as a strictly decreasing extension of $p_i R'_i$ over the whole real axis, i.e., $\tilde{R}''_i(\tau)=p_i R'_i(\tau)$ for all $\tau\in[0;1]$ and $\tilde{R}''_i$ is strictly decreasing over $\mathbb R$. 

\noindent\textbf{Step 2} Compute $\mu^{(k)}$ via the equation $\sum_{i\in \mc M^{(k)}} S_i [\tilde{R}''_i]^{-1}(\mu^{(k)})=C^{(k)}$.\\
Compute the sets $\mc M_a^{\mu(k)}=\{ m: \, [\tilde{R}''_i]^{-1}(\mu^{(k)})<0 \}$, $\mc M_b^{\mu(k)}=\{ m: \, [\tilde{R}''_i]^{-1}(\mu^{(k)})>1 \}$, $\mc M^{\mu(k)}=\{ m: \, 0\le [\tilde{R}''_i]^{-1}(\mu^{(k)})\le 1 \}$.\\
Compute $\delta ( \mu^{(k)} ) = \, \sum_{i\in \mc M_b^{\mu(k)}} S_i + \sum_{i\in \mc M^{\mu(k)}} S_i [R'_i]^{-1}(\mu^{(k)}) - C^{(k)}$.

\noindent\textbf{Step 3} If $\delta (\mu^{(k)})=0$ or $\mc M^{\mu(k)}=\emptyset$ then set $\mu=\mu(k)$, $\mc M_a^{\mu}=\mc M_a^{\mu}\cup\mc M_a^{\mu(k)}$, $\mc M_b^{\mu}=\mc M_b^{\mu}\cup\mc M_b^{\mu(k)}$, $\mc M^{\mu}=\mc M^{\mu(k)}$, and go to step 6.\\
Else, if $\delta (\mu^{(k)})>0$ then go to step 4.\\
Else, if $\delta (\mu^{(k)})<0$ then go to step 5.

\noindent\textbf{Step 4} Set $\eta_i^*=0$ for all $i\in \mc M_a^{\mu(k)}$. Set $C^{(k+1)}:=C^{(k)}$. Compute $\mc M^{(k+1)} := \mc M^{(k)}\setminus \mc M_a^{\mu(k)}$, $\mc M_a^{\mu} := \mc M_a^{\mu} \cup\mc M_a^{\mu(k)}$, $k:=k+1$. Go to step 2.

\noindent\textbf{Step 5} Set $\eta_i^*=S_i$ for all $i\in \mc M_b^{\mu(k)}$. Compute $C^{(k+1)}=C^{(k)}-\sum_{i\in \mc M_b^{\mu(k)}} S_i$, $\mc M^{(k+1)} := \mc M^{(k)}\setminus \mc M_b^{\mu(k)}$, $\mc M_b^{\mu} := \mc M_b^{\mu} \cup\mc M_b^{\mu(k)}$, $k:=k+1$. Go to step 2.

\noindent\textbf{Step 6} Set $\eta_i^*=0$ for all $i\in \mc M_a^{\mu}$; $\eta_i^*=1$ for all $i\in \mc M_b^{\mu}$; $\eta_i^*=[\tilde{R}''_i]^{-1}(\mu^{(k)})$ for all $i\in \mc M^{\mu}$. Stop.

\noindent\rule{3.5in}{0.01in}
\end{footnotesize}

\subsection{Proof of Theorem \ref{theo:optimalwater}}

\begin{proof}
As a first step, let us define $f_i(\tau): \ [0;1]\rightarrow [0;1]$ as a one-to-one function such that the permuted audience retention rate function $R_i'(\tau) := R_i(f^{-1}_i(\tau))$ is non decreasing. The function $f_i$ is a permutation function that orders the video parts in order of decreasing popularity, such that $f_i(\tau)<f_i(\tau')$ if and only if $R_i(\tau)>R_i(\tau')$\footnote{We notice that such $f_i$ always exists, even though is not unique, since it can arbitrarily break the ties among equally popular parts of a single video, and it is in general discontinuous.}. Then, $R'_i$ is the outcome of such permutation. As a second step, we reformulate the optimization problem in \eqref{eq:uppbound1} as
\begin{align}
\boldsymbol{Y}^* = \underset{\boldsymbol Y}{\operatorname{argmax}} & \, \sum_{i\in\mc M} S_i \int_{Y_i} p_i R_i(\tau) d\tau \label{eq:uppbound22} \\
\mathrm{s.t.} & \, \begin{sistema}
\sum_{i\in\mc M} S_i \int_{Y_i} 1 d\tau = \, C \\
Y_i\subseteq [0;1]
\end{sistema} \notag
\end{align}
We can recast the bandwidth saving optimization problem in \eqref{eq:uppbound22} in terms of the permuted engagement rates $R'_i$ and by considering only right intervals of 0 of the kind $Y_i=[0;\eta_i]$, as follows:
\begin{align}
\max_{\boldsymbol \eta \in\mathbb{R}^M} & \, \sum_{i\in\mc M} p_i S_i \int_0^{\eta_i} R'_i(\tau) d\tau \label{eq:uppbound2} \\
\mathrm{s.t.} & \, \begin{sistema}
\sum_{i\in\mc M} \eta_i S_i = \, C \\
\eta_i \in [0;1].
\end{sistema} \notag
\end{align}
In fact, it is not profitable to consider a larger search domain, e.g., more complicated subsets $\boldsymbol{Y}$ of $[0;1]^M$: for any collection of subsets $\boldsymbol{Y}$ it is possible to replace $Y_i$ with the interval $[0;\int_{Y_i}d\tau]$ with a strict increase of the objective function while the feasibility is still preserved. We can further simplify \eqref{eq:uppbound2} by defining the function $R''_i(\tau)=p_i R'_i(\tau)$, as follows:
\begin{align}
\min_{\boldsymbol \eta \in\mathbb{R}^M} & \, \sum_{i\in\mc M} \int_0^{\eta_i} -R''_i(\tau) d\tau \label{eq:uppbound3} \\
\mathrm{s.t.} & \, \begin{sistema}
\sum_{i\in\mc M} \eta_i = \, C \\
\eta_i S_i \in \left[0;S_i\right].
\end{sistema} \notag
\end{align}
We notice that $\frac{d}{d\eta_i}\int_0^{\eta_i} -R''_i(\tau) d\tau=-p_i R'_i(\eta_i)$, which is non-decreasing in $\eta_i$. Thus we recognize in \eqref{eq:uppbound3} a convex optimization problem with linear and box constraints, where the objective function is separable in the optimization variables $\boldsymbol{\eta}$. It is known that such kind of problems can be solved via a classic water-filling technique (see \cite{stefanov2013separable}, Chapter 6): more specifically, there exists a positive ``water level'' $\mu$ such that the optimal portions $\boldsymbol{\eta}^*(\mu)$ can be computed as
\begin{equation} \label{eq:uppbound}
\begin{sistema}
{\eta_i}^*(\mu) = \, \begin{sistema}
1 \quad \mathrm{if} \ \min_{\tau\in [0;1]} R_i''(\tau)\ge \mu \\
0 \quad \mathrm{if} \ \max_{\tau\in [0;1]} R_i''(\tau)\le \mu \\
R_i^{''-1}(\mu) \quad \mathrm{else} 
\end{sistema} \\
\sum_{i\in\mc M} S_i \eta_i^*(\mu) = \, C
\end{sistema}
\end{equation}
By rewriting \eqref{eq:uppbound} in terms of $R'_i$ we obtain the expressions:
\[
\begin{sistema}
\eta_i^* = \, \begin{sistema}
1 \quad \mathrm{if} \ p_i \min_{\tau\in [0;1]} R'_i(\tau)\ge \mu \\
0 \quad \mathrm{if} \ p_i \max_{\tau\in [0;1]} R'_i(\tau) \le \mu \\
R_i^{'-1}(\mu/p_i) \quad \mathrm{else} 
\end{sistema} \\
\sum_{i\in\mc M} S_i |Y_i^*| = \, C.
\end{sistema}
\]
and we can finally claim that 
\[
Y_i^* = \, f_i^{-1}\left(\left[0;\eta_i^* \right]\right) = \, \{\tau: \ p_i R_i(\tau)\ge \mu \} \quad \forall\, i\in\mc M.
\]
The thesis follows.
\end{proof}

\subsection{Proof of Proposition \ref{prop:downlwater}}

\begin{proof}
Since $R_i$ is already strictly decreasing, then we can consider $f_i(\tau)=\tau$ and $R'_i=R_i$. Moreover, in this case $\min_{\tau} R_i(\tau)=0$ and $\max_{\tau} R_i(\tau)=1$. The thesis easily follows. 
\end{proof}

\subsection{Proof of Corollary \ref{cor:downlexpwater}}

\begin{proof}
Define 
\[
\tilde{R}^{-1}_i(\tau) = \, -\frac{1}{\lambda_i} \ln\left( \tau(1-e^{-\lambda_i}) + e^{-\lambda_i}\right).
\]
We notice that $\tilde{R}^{-1}_i(\mu/p_i)=R_i^{-1}(\mu/p_i)$ when $0<\mu\le p_i$ and $\tilde{R}^{-1}_i(\mu/p_i)<0$ whenever $p_i>\mu$. Then, we can rewrite \eqref{eq:cor1} as 
\[
\begin{sistema}
\eta_i^*= \, \left[ \tilde{R}_i^{-1}(\mu/p_i) \right]^+ \\
\sum_{i\in\mc M} S_i \eta_i^* = \, C.
\end{sistema}
\]
The thesis easily follows.
\end{proof}

\subsection{Proof of Theorem \ref{theo:infchunk}}

\begin{proof}
Let us first introduce the function 
\[
\xi^{(t_C)} (\tau) = \, \sum_{i=1}^M p_i R_i(\tau) e^{-p_i R_i(\tau) t_C}. 
\]
We then define $\mathcal I(f)|_{\mathbf{x}}$, where $f$ is a continuous function defined over $\mathbb R$, the integral approximation of $f$ via Riemann sums of the type:
\[
\mathcal I(f)|_{\mathbf{x}} = \, \sum_{k=1}^{N} f(x_{k-1}) \Delta x_k.
\]
We notice that if $f$ is increasing (decreasing) then $\mathcal I(f)|_{\mathbf{x}}<(>)\mathcal I(f)|_{\mathbf{x}'}$ for any sub-splitting $\mathbf{x}'$. We can now rewrite $B_{\mathrm{cLRU}} (\mathbf{x},\nu)$ as (compare with \eqref{eq:BcLRU})
\begin{align*}
B_{\mathrm{cLRU}} (\mathbf{x},\nu) = & \, \mathcal{I} (\xi^{(t_C)})|_{\mathbf{x}} \\
\mathrm{s.t.} \ M\nu - \frac{C}{S} = & \, \mathcal{I} (h^{(t_C)})|_{\mathbf{x}}
\end{align*}
where $h^{(t_C)} (\tau)=\sum_{i=1}^M e^{-p_i R_i(\tau) t_C}$. Since $h^{(t_C)} (\tau)$ is increasing in $\tau$, it easily follows from an induction argument that the value of characteristic time for any chunk splitting is found within $[\underline{t}_C;\overline{t}_C]$.\\
Consider now a sub-splitting $\mathbf{x}'$ with associated characteristic time $t'_C$. Since $h^{(t_C)} (\tau)$ is increasing, then $\mathcal{I} (h^{(t_C)})|_{\mathbf{x}'}>\mathcal{I} (h^{(t_C)})|_{\mathbf{x}}$. Also, since $\mathcal{I} (h^{(t'_C)})|_{\mathbf{x}'}=\mathcal{I} (h^{(t_C)})|_{\mathbf{x}}$, and $h^{(t)} (\tau)$ is decreasing in $t$ then $t'_C>t_C$. We then have 
\begin{align*}
B_{\mathrm{cLRU}} (\mathbf{x},\nu) = \mathcal{I} (\xi^{(t_C)})|_{\mathbf{x}} > \mathcal{I} (\xi^{(t'_C)})|_{\mathbf{x}} > & \, \mathcal{I} (\xi^{(t'_C)})|_{\mathbf{x}'} \\ 
= & \, B_{\mathrm{cLRU}} (\mathbf{x}',\nu)
\end{align*}
where the second inequality follows from the fact that $\xi^{(t)} (\tau)$ is decreasing in $\tau$ for any  value $t$ of the characteristic time. The thesis is proven.
\end{proof}

\subsection{Proof of Corollary \ref{cor:nuoptmin1}}

\begin{proof}
The derivative with respect to $\nu$ of the objective function in \eqref{eq:infinitesimalLRU} in the direction along which the constraint is satisfied writes
\begin{align}
& q(\nu) = - \sum_{i=1}^M (1-e^{-p_i R_i(\nu) t_C})p_i R_i(\nu) + \label{eq:derder} \\
& \int_0^{\nu} \sum_{i=1}^M p_i^2 R_i^2(\tau) e^{-p_i R_i(\tau) t_C} d\tau \frac{\sum_{i=1}^M 1-e^{-p_i R_i(\nu) t_C}}{\int_0^{\nu} \sum_{i=1}^M p_i R_i(\tau) e^{-p_i R_i(\tau) t_C} d\tau} \notag 
\end{align}
Let us calculate $q(1-d\nu)$, which equals
\begin{align*}
& \, d\nu \left( \frac{A+B \, d\nu }{C+D\, d\nu}\, \sum_{i=1}^M p_i |R'_i(1)| - d\nu \sum_{i=1}^M p_i^2 |R'_i(1)|^2 \right).
\end{align*}
Since $A= \int_0^{\nu} \sum_{i=1}^M p_i^2 R_i^2(\tau) e^{-p_i R_i(\tau) t_C} d\tau>0$ and $B=\int_0^{\nu} \sum_{i=1}^M p_i R_i(\tau) e^{-p_i R_i(\tau) t_C} d\tau>0$, then $q(1-d\nu)>0$ and thesis is proven.
\end{proof}

\subsection{Proof of Corollary \ref{cor:nuopteq1}}

\begin{proof}
We first observe that, if $R_i(\tau)=1$, then for all $\nu$ we have $B_{\mathrm{cLRU}}([0;\nu],\nu)=B_{\mathrm{cLRU}}(\mathbf{x},\nu)$ for any chunk splitting $\mathbf x$. Then it suffices to prove that $q(\nu)<0$ holds for all $\nu\in(0;1)$, i.e., that the following expression holds:
\begin{align*}
& \left(\sum_{i=1}^M 1-e^{-p_i t_C}\right)\sum_{i=1}^M p_i^2 e^{-p_i t_C} + \\
& - \sum_{i=1}^M (1-e^{-p_i t_C})p_i \sum_{i=1}^M p_i e^{-p_i t_C} <0
\end{align*}

\end{proof}


\begin{thebibliography}{10}
\providecommand{\url}[1]{#1}
\csname url@samestyle\endcsname
\providecommand{\newblock}{\relax}
\providecommand{\bibinfo}[2]{#2}
\providecommand{\BIBentrySTDinterwordspacing}{\spaceskip=0pt\relax}
\providecommand{\BIBentryALTinterwordstretchfactor}{4}
\providecommand{\BIBentryALTinterwordspacing}{\spaceskip=\fontdimen2\font plus
\BIBentryALTinterwordstretchfactor\fontdimen3\font minus
  \fontdimen4\font\relax}
\providecommand{\BIBforeignlanguage}[2]{{%
\expandafter\ifx\csname l@#1\endcsname\relax
\typeout{** WARNING: IEEEtran.bst: No hyphenation pattern has been}%
\typeout{** loaded for the language `#1'. Using the pattern for}%
\typeout{** the default language instead.}%
\else
\language=\csname l@#1\endcsname
\fi
#2}}
\providecommand{\BIBdecl}{\relax}
\BIBdecl

\bibitem{roberts2013exploring}
J.~Roberts and N.~Sbihi, ``Exploring the memory-bandwidth tradeoff in an
  information-centric network,'' in \emph{Proc. of ITC}, 2013, pp. 1--9.

\bibitem{cisco_VNI}
``Cisco visual networking index: Forecast and methodology, 2014–2019,''
  \url{http://www.cisco.com/c/en/us/solutions/collateral/service-provider/ip-ngn-ip-next-generation-network/white_paper_c11-481360.html}.

\bibitem{hwang12}
K.~W. Hwang, D.~Applegate, A.~Archer, V.~Gopalakrishnan, S.~Lee, V.~Misra,
  K.~K. Ramakrishnan, and D.~F. Swayne, ``Leveraging video viewing patterns for
  optimal content placement,'' in \emph{Proceedings of IFIP Conference on
  Networking}, ser. IFIP'12, 2012, pp. 44--58.

\bibitem{retentionrate}
\url{http://support.google.com/youtube/answer/1715160?hl=en-GB}.

\bibitem{zeni2013youstatanalyzer}
M.~Zeni, D.~Miorandi, and F.~De~Pellegrini, ``{YOUStatAnalyzer}: a tool for
  analysing the dynamics of {YouTube} content popularity,'' in \emph{Proc of
  VALUETOOLS 13}.\hskip 1em plus 0.5em minus 0.4em\relax ICST, 2013, pp.
  286--289.

\bibitem{Sen99}
S.~Sen, J.~Rexford, and D.~Towsley, ``Proxy prefix caching for multimedia
  streams,'' in \emph{Proc. of IEEE INFOCOM '99}, vol.~3, Mar 1999, pp.
  1310--1319 vol.3.

\bibitem{wu04}
K.-L. Wu, P.~Yu, and J.~Wolf, ``Segmentation of multimedia streams for proxy
  caching,'' \emph{IEEE Transactions on Multimedia}, vol.~6, no.~5, pp.
  770--780, Oct 2004.

\bibitem{wang2015optimal}
L.~Wang, S.~Bayhan, and J.~Kangasharju, ``Optimal chunking and partial caching
  in information-centric networks,'' \emph{Computer Communications}, vol.~61,
  pp. 48--57, 2015.

\bibitem{yu2006dynamic}
J.~Yu, C.~T. Chou, Z.~Yang, X.~Du, and T.~Wang, ``A dynamic caching algorithm
  based on internal popularity distribution of streaming media,''
  \emph{Multimedia Systems}, vol.~12, no.~2, pp. 135--149, 2006.

\bibitem{agrawal14}
K.~Agrawal, T.~Venkatesh, and D.~Medhi, ``A dynamic popularity-based partial
  caching scheme for video on demand service in {IPTV} networks,'' in
  \emph{Proc. of COMSNETS ' 14}, Jan 2014, pp. 1--8.

\bibitem{lim14}
S.-H. Lim, Y.-B. Ko, G.-H. Jung, J.~Kim, and M.-W. Jang, ``Inter-chunk
  popularity-based edge-first caching in content-centric networking,''
  \emph{IEEE Communications Letters}, vol.~18, no.~8, pp. 1331--1334, Aug 2014.

\bibitem{chen05}
S.~Chen, H.~Wang, X.~Zhang, B.~Shen, and S.~Wee, ``Segment-based proxy caching
  for {Internet} streaming media delivery,'' \emph{IEEE Multimedia}, vol.~12,
  no.~3, pp. 59--67, 2005.

\bibitem{hefeeda08}
M.~Hefeeda and O.~Saleh, ``Traffic modeling and proportional partial caching
  for peer-to-peer systems,'' \emph{IEEE/ACM Transactions on Networking,},
  vol.~16, no.~6, pp. 1447--1460, Dec 2008.

\bibitem{devi12}
U.~Devi, R.~Polavarapu, M.~Chetlur, and S.~Kalyanaraman, ``On the partial
  caching of streaming video,'' in \emph{IEEE IWQoS, 2012}, June 2012, pp.
  1--9.

\bibitem{che2002hierarchical}
H.~Che, Y.~Tung, and Z.~Wang, ``Hierarchical web caching systems: Modeling,
  design and experimental results,'' \emph{IEEE Journal on Selected Areas in
  Communications,}, vol.~20, no.~7, pp. 1305--1314, 2002.

\bibitem{cleveland1979robust}
W.~S. Cleveland, ``Robust locally weighted regression and smoothing
  scatterplots,'' \emph{Journal of the American statistical association},
  vol.~74, no. 368, pp. 829--836, 1979.

\bibitem{fricker2012versatile}
C.~Fricker, P.~Robert, and J.~Roberts, ``A versatile and accurate approximation
  for lru cache performance,'' in \emph{24th International Teletraffic Congress
  (ITC 24)}, Sept 2012, pp. 1--8.

\bibitem{wistiaeng}
\url{http://wistia.com/doc/audience-engagement-graph}.

\bibitem{stefanov2013separable}
S.~M. Stefanov, \emph{Separable programming: theory and methods}.\hskip 1em
  plus 0.5em minus 0.4em\relax Springer Science \& Business Media, 2013,
  vol.~53.

\end{thebibliography}


\end{document}